\documentclass[letterpaper, 10 pt, conference]{ieeeconf}  

\IEEEoverridecommandlockouts                              

\usepackage{amssymb,amsfonts,amsmath,amsthm,amscd}
\usepackage{epsfig,graphics,psfrag}
\usepackage{url}

\newtheorem{propo}{Proposition}[section]
\newtheorem{lemma}[propo]{Lemma}

\newtheorem{thm}[propo]{Theorem}

\newcommand{\reals}{{\mathds R}}

\newcommand{\eqnsection}{\renewcommand{\theequation}{\thesection.\arabic{equation}}
      \makeatletter \csname @addtoreset\endcsname{equation}{section}\makeatother}
\def\eps{\epsilon}
\def\l|{\left|\left|}
\def\r|{\right|\right|}

\def\E{\mathbb E}
\def\M{{\sf M}}
\def\U{{\sf U}}
\def\V{{\sf V}}
\def\prob{{\mathbb P}}
\def\ed{\stackrel{\rm d}{=}}
\def\ind{{\mathbb I}}
\def\ve{\varepsilon}
\def\te{\widetilde{\eps}}

\def\reals{{\mathbb R}}

\def\vv{\vec{v}}
\def\vu{\vec{u}}
\def\l0{p_0}
\def\r0{q_0}

\def\hM{\widehat{{\sf M}}}

\def\odelta{\overline{\delta}}
\def\sdelta{\overline{\delta}^u}
\def\dmax{\overline{d}}
\def\cD{{\cal D}}
\def\cC{{\cal C}}
\def\Typ{{\sf Typ}}
\def\Zh{\widehat{Z}}

\renewcommand{\theenumi}{{\roman{enumi}}}

\title{\Large \bf Learning Low Rank Matrices from $O(n)$ Entries}

\author{Raghunandan H. Keshavan, Andrea Montanari and Sewoong Oh
\thanks{Raghunandan H. Keshavan 
is with the Department of Electrical Engineering,
Stanford University, {\tt\small raghuram@stanford.edu}. Andrea Montanari is with Departments
of Electrical Engineering and Statistics, Stanford University, {\tt\small
montanari@stanford.edu}. Sewoong Oh is with the Department of Electrical
Engineering, Stanford University, {\tt\small swoh@stanford.edu}.} }

\begin{document}

\maketitle

\begin{abstract}
How many random entries of an $n\times n\alpha$, rank $r$ matrix are necessary
to reconstruct the matrix within an accuracy $\delta$? We address
this question in the case of  a random matrix  with bounded
rank, whereby the observed entries are chosen uniformly at random.
We prove that, for any $\delta>0$, $C(r,\delta)n$ observations are
sufficient.

Finally we discuss the question of reconstructing the matrix
\emph{efficiently}, and demonstrate through extensive simulations 
that this task can be accomplished in $n$Poly$(\log n)$ operations, 
for small rank. 
\end{abstract}
%
%
\section{Introduction and main results}

\subsection{Problem definition}

Let $\M$ be an $n\times m$ matrix of rank (at most) $r$
and assume that $n\eps$ uniformly random entries of $\M$ 
are revealed. Does this knowledge allow to approximately 
reconstruct $\M$? 

The answer is negative unless the matrix has some specific structure.
In this paper we assume that $\M$ is a \emph{random rank-$r$ matrix},
i.e. $\M = \U\cdot\V$ where $\U$ is a $n\times r$ matrix with iid
entries and $\V$ an independent $r\times m$ matrix with iid entries.
The distributions of the entries of $\U$ and $\V$ are denoted, 
respectively as $p_0$ and $q_0$.

The metric we shall consider is the root mean square error (RMSE). If 
$\{\M_{i,a}\}$ are the entries of $\M$, and $\hM$ is its estimate based
on the observed entries, we have
\begin{eqnarray}
D(\M,\hM) \equiv
\Big\{\frac{1}{nm}\sum_{i,a} |\M_{i,a}-\hM_{i,a}|^2\Big\}^{1/2}\, .
\label{eq:RMSEDef}
\end{eqnarray}
Notice that this coincides, up to a factor, with the distance induced
by the Frobenius norm $D(\M,\hM) = ||\M-\hM||_{\rm F}/\sqrt{nm}$.

In the following we shall denote by $R\ni i,j,k,\dots$ the set of rows of 
$\M$ and by $C\ni a,b,c,\dots$ its set of columns. The subset
of revealed entries will be denoted by $E \subseteq R\times C$.
%
%
\subsection{Motivation and related work}

Low rank matrices have been proposed as statistical models 
to describe a number of complex data sources. For instance,
the matrix of empirical correlations among stock prices in a market
is approximately low rank if price fluctuations are driven by 
a few underlying mechanisms \cite{Johnstone}. 
A completely different application 
is provided by the matrix of square distances among $n$ sensors 
in $3$ dimension, which has rank $r=5$ \cite{Sensors}.

Low rank matrices have been proposed as a model
for collaborative filtering data. As a concrete example we
shall focus here on the Netflix Challenge dataset \cite{Netflix}.
This dataset concerns a set $C$ of approximately $5\cdot 10^5$
customers and $R$ of $2\cdot 10^4$ movies. For
about $10^8$ customer-movie pairs $(i,a)\in E$, the corresponding
rating (an integer between $1$ and $5$) is provided.
The challenge consists in predicting the ratings of $10^6$ 
non-revealed customer-movie pairs within a root mean square 
error smaller than $0.8563$.

One possible approach consists in considering the customer-movie matrix
$\M$ (or a rescaled version of it) and assuming that it has low rank 
to predict the requested entries. Indeed, a simple coordinate descent 
algorithm that minimizes the energy function
\begin{eqnarray}
\sum_{(i,a)\in E}(\M_{i,a}-(\U\V)_{i,a})^2+\lambda ||\U||^2_{\rm F}
+\lambda ||\V||^2_{\rm F}\label{eq:Regularized}
\end{eqnarray}
provides good predictions (within the Netflix competition,
it was used by SimonFunk).

In general, the matrix completion problem is not convex, and the descent 
algorithm is not guaranteed to converge to the original matrix $\M$
even if this is the unique rank $r$ matrix consistent with the observations.
A possible alternative consists in relaxing the rank constraint,
by looking instead for a matrix $\hM$ of minimal nuclear norm 
(recall that the nuclear norm of $\hM$ is the sum of the absolute
values of its singular values).
The problem then becomes convex and indeed reducible to semidefinite 
programming. In \cite{Parrillo} it was shown that this relaxation indeed
recovers the original low rank matrix $\M$, given that
a sufficient number of random 
linear combinations of its entries are revealed.

The case in which a random subset of the entries is
revealed (which is relevant for
collaborative filtering) was treated in \cite{Candes}.
This paper proves that the convex relaxation is tight with high
probability\footnote{Strictly speaking, the matrix model treated 
in \cite{Candes} is slightly different from the one considered here.
However it should not be hard to prove that the two models
are asymptotically equivalent for large $n$.} 
if $\eps \ge C\, r\, n^{1/5}\log n$.
In particular this implies two statements:
$(i)$ For $\eps \ge C\, r\, n^{1/5}\log n$, $n\eps$ random
entries uniquely determine a random rank-$r$ matrix.
$(ii)$ This matrix is the unique minimum of a semidefinite program.
%
%
\subsection{Main results}

The results briefly reviewed above leave open several key issues:
\begin{enumerate}
\item[1.] Why is it necessary to observe $\Theta(n^{6/5})$ entries
to reconstruct a rank-$r$ matrix, that has $\Theta(n)$ degrees of freedom?
\item[2.] As the Netflix challenge shows, it is not realistic nor necessary to 
reconstruct $\M$ exactly. What is the trade-off between 
RMSE distortion and number of observations?
\item[3.] In general, semidefinite programming has $\Theta(n^6)$
complexity \cite{SDP}.
This is affordable up to $n\approx 10^2$, but way beyond current
capabilities when $n\approx 10^5$ as in modern datasets. 
\end{enumerate} 

In this paper we address the first two points and show that
$O(n)$ observations are sufficient to reconstruct a low 
rank matrix within any positive distortion. 
\begin{thm}\label{thm:MainUB}
Let $\M = \U\cdot \V$ be a random rank-$r$ matrix with $n$ rows and
$n\alpha$ columns and assume the distributions of $\U_{i,k}$ and $\V_{k,a}$ 
to have support in $[-1,1]$. Let $E$ be a random subset of $n\eps$ 
entries in $R\times C$.
Then, with high probability, any rank-$r$ matrix $\hM$ such that 
$|\M_{i,a}-\hM_{i,a}|\le \Delta$ for all $(i,a)\in E$, 
and with factors  $\U_{i,k},\V_{k,a}\in[-1,1]$, also satisfies
\begin{eqnarray}
D(\M,\hM)\le \Delta + 2r\; \te^{-1/2}\log(10\te)\, ,
\end{eqnarray}
where $\te \equiv \eps/(1+\alpha)r$.
\end{thm}
Notice that the term $\Delta$ in the above inequality is unavoidable.
Since we are looking for matrices that match the observed entries 
only within precision $\Delta$, we cannot hope for a RMSE
smaller than $\Delta$. In the second term, the factor $2r$ corresponds to
the maximal distance between matrix entries in the present model, while
the $\eps$-dependent factor tends to $0$ as $\eps\to\infty$.
Notice that $\te$ is exactly the number of observations per
degree of freedom.

The proof of this statement is given in Section \ref{sec:UB},
which also provides a much more accurate upper bound. The latter is
--however--
not straightforward to evaluate. While it is clear that 
small RMSE cannot be achieved with less than $\Theta(n)$ observed 
matrix elements, Section \ref{sec:LB} proves a quantitative lower bound of 
this form.

In Section \ref{sec:Numerical} we address the question of efficient 
reconstruction and demonstrate that $O(n\log n)$ operations are 
sufficient to reconstruct random low rank matrices with rank $r\le 4$,
from $O(n)$ entries.
Indeed such performances are achieved by a straightforward stochastic
local search algorithm that we refer to as WalkRank or by a coordinate descent
algorithm. A formal analysis of these algorithms will be presented in a future 
publication. 
Finally, in Section \ref{sec:Netflix} we use these results to
compare random low rank matrices and the Netflix dataset.

Before dwelling on the intricacies of the full problem,
the next Section discusses a particularly simple but perhaps
instructive case: rank $r=1$.
%
%
\section{A warmup example}
\label{sec:Warmup}

If $\M$ has rank $1$, most of the questions listed above have a simple
answer with a suggestive graph-theoretical interpretation.

Assume that you know $3$ entries of the matrix $\M$ that belong to the 
same $2\times 2$ minor. Explicitly, for two row indices $i,j\in R$
and two column indices $a,b\in C$, the entries $\M_{i,a}$,
$\M_{j,a}$, $\M_{i,b}$ are known. Unless $\M_{i,a}=0$,
the fourth entry of the same minor is then uniquely 
determined $\M_{j,b} = \M_{j,a}\M_{i,b}/\M_{i,a}$. The case $\M_{i,a}=0$
can be treated separately but, for the sake of simplicity we 
shall assume that the distributions $p_0$, $q_0$ do not have mass on $0$.

This observation suggests a simple matrix completion algorithm: 
Recursively look for a $2\times 2$ minor with a unique unknown entry
and complete it according to the rule $\M_{j,b} = \M_{j,a}\M_{i,b}/\M_{i,a}$.
As anticipated above, this algorithm has a nice graph-theoretic
interpretation. Consider the bipartite graph $G = (R,C,E)$ with 
vertices corresponding to the row and columns of $\M$ and edges 
for the observed entries. If a $2\times 2$ minor has a unique unknown entry,
it means that the corresponding vertices $j\in R$, $b\in C$ are connected 
by a length-$3$ path in $G$. Hence the algorithm recursively adds 
edges to $G$ connecting distance-$3$ vertices.

\begin{figure}[h]
\center{\includegraphics[width=8.cm]{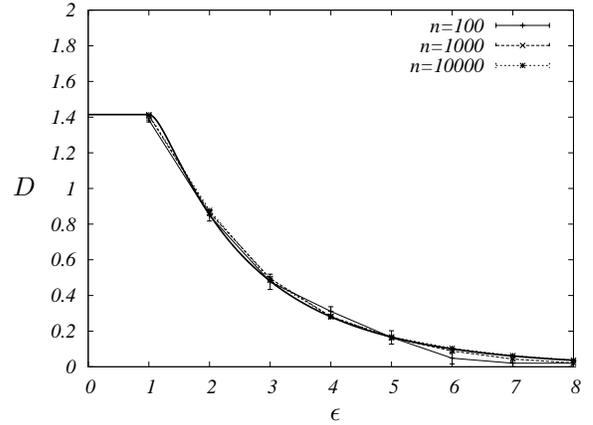}}
\put(-105,-5){$\eps$}
\put(-225,80){$D$}
\caption{{\small Learning random rank-$1$ matrices. The continuous 
line is the optimal distortion (achieved by the recursive completion algorithm).
Data points correspond to a $O(n)$ complexity local search algorithm.}}
\label{fig:Rank1Walk}
\end{figure}
After at most $O(n^2)$ operations the process described halts on a graph that 
is a disjoint union of cliques, corresponding to the connected
components in $G$. Each edge corresponds to a correctly predicted matrix
entry. Clearly, in the large $n$-limit only the components with $\Theta(n)$
matter (as they have $\Theta(n^2)$ edges). It is a fundamental result 
in random graph theory that there is no such component for 
$\eps\le 1/\sqrt{\alpha}$. For $\eps>1/\sqrt{\alpha}$ there is one such 
component involving approximately $n\xi$ in $R$ and $m\zeta$ vertices in $C$,
where $(\xi,\zeta)$ is the unique positive  solution of
\begin{eqnarray}
\xi = 1-e^{-\eps\alpha \zeta}\, ,\;\;\;\;\;\;\; \zeta = 1-e^{-\eps\xi}\, .
\label{eq:GiantComponent}
\end{eqnarray}

This analysis implies the following result.
\begin{propo}\label{propo:Rank1}
Let $\M = \U\cdot \V$ be a random rank $1$ matrix, and denote
by $\xi(\eps)$, $\zeta(\eps)$ the largest solution of 
Eq.~(\ref{eq:GiantComponent}). Then there exists an algorithm
with $O(n^2)$ complexity achieving, with high probability, RMSE
\begin{eqnarray}
D(\M,\hM) = \sqrt{1-\xi(\eps)\zeta(\eps)}\, D_0+O(\sqrt{(\log n)/n})\, .
\end{eqnarray}
where $D_0\equiv\sqrt{\E(V_1^2)\E(U_1^2)}$. 
Further, if the entries $\U_i$, $\V_a$ have symmetric distribution,
then no algorithm achieves smaller distortion.
\end{propo}
\begin{proof}
The mentioned distortion is achieved by the recursive completion algorithm,
whereby matrix element corresponding to vertex pairs in distinct components
are predicted to vanish. This is optimal if the matrix element 
distribution is symmetric. Indeed the conditional matrix element 
distribution remains symmetric even given the observations. 
\end{proof}

For massive datasets even $O(n^2)$ complexity is unaffordable.
Figure \ref{fig:Rank1Walk} compares the minimal distortion
guaranteed by Proposition \ref{propo:Rank1} with the performances of
the WalkRank algorithm described in Section \ref{sec:Numerical}.
Here the factors $\U_i$, $\V_a$ where chosen uniformly in $\{+1,-1\}$.
%
%
\section{Upper Bound and Proof of Theorem \ref{thm:MainUB}}
\label{sec:UB}

In this section we prove the upper bound on distortion
stated in Theorem \ref{thm:MainUB}. The proof proceeds in three steps.
First we will consider the case in which the factor entries 
$\U_{i,k}$, $\V_{k,a}$ are supported on a finite set, and prove a (tighter)
upper bound via a counting argument. Then we'll use a quantization argument
to generalize this bound to the continuous case. Finally, we simplify our 
bound to get the pleasing expression in Theorem \ref{thm:MainUB}.
Unfortunately this simplification entails a worsening of the bound.
%
%
%
\subsection{The discrete case}
\label{subsec:discrete}
We start by introducing a couple of new notations.
Given a row index $i\in R$, we let $\vu^0_i=(\U_{i,1},\dots,\U_{i,r})$ 
be the $i$-th row of $\U$. Analogously, for $a\in C$, let $\vv^0_a$ be
the $a$-th column of $\V$. We then have
\begin{eqnarray}
\M_{i,a} = \vu^0_i\cdot \vv^0_a\, .\label{eq:MatrixForm}
\end{eqnarray}
We also write $\vu^0_i = (u^0_{i,1},\dots,u^0_{i,r})$
and  $\vv^0_a = (v^0_{a,1},\dots,v^0_{a,r})$ for the components 
of these vectors. These are assumed to be iid's with distributions 
$p_0$ (for $\vu$) and $q_0$ (for $\vv$) supported on a
finite set $A_N\subset \reals$ with $|A_N| = N$ points.
Typical examples are $A_2 = \{-1,+1\}$ or 
$A_{2M+1}\equiv \{-M\ve,-(M-1)\ve,\dots,(M-1)\ve,M\ve\}$).
Our basic counting estimate is stated below.
\begin{propo}
\label{propo:discrete}
Let $\Delta\ge 0$ and $\M$ be a random rank-$r$ matrix
with factors supported in $A_N$. Then, with high probability any 
rank-$r$ matrix $\hM$ with factors supported in $A_N$ 
that satisfies $|\M_{ia}-\hM_{ia}|\le \Delta$ for all
$(i,a)\in E$ also satisfies $D(\M,\hM)\le \odelta(\eps,\alpha,\Delta) 
+o_n(1)$, where
\begin{eqnarray}
\odelta(\eps,\alpha,\Delta) = \!\!\sup_{p\in\cD(p_0),q\in\cD(q_0)}\!\!\!
\left\{\;d(p,q)\,:\;\; \phi_{\Delta}
(p,q)\;\ge 0\right\}\, .\label{eq:Upperbound}
\end{eqnarray}
Here the $\sup$ over $p$ (over $q$) is taken over the space  of distributions
$\cD(p_0)$ (respectively $\cD(q_0)$) over $(A_N)^r\times (A_N)^r$ 
such that $\sum_{\vu}p(\vu,\vu^0) = p_0(\vu^0)$ (respectively
$\sum_{\vv}q(\vv,\vv^0) = q_0(\vv^0)$).
The functionals appearing in Eq.~(\ref{eq:Upperbound}) are defined by
\begin{eqnarray}
d(p,q) \equiv \left\{\E_{p,q}\; |\vu\cdot\vv-\vu^0\cdot\vv^0|^2 
\right\}^{1/2}\, ,
\end{eqnarray}
and
\begin{align}
\phi_\Delta(p,q) &\equiv H(p)-H(p_0)+\alpha[H(q)-H(q_0)]+\\
&+\eps\, \E_{p_0,q_0}\log
\prob_{p,q}\big\{|\vu\cdot\vv-\vu^0\cdot\vv^0| \le \Delta\;\big|\;\vu^0,\vv^0\big\}\, ,\nonumber
\end{align}
\end{propo}
\begin{proof}

 Define $Z_G(\Delta,\delta)$ 
($G$ is the bipartite graph
with edge set $E$) as the number of matrices $\hM$ of the form
(\ref{eq:MatrixForm}) such that:
\begin{enumerate}
\item[$(1)$] $|\M_{i,a}-\hM_{i,a}|\le \Delta$ for all $(i,a)\in E$;
\item[$(2)$] $D(\M,\hM)\ge \delta$.
\end{enumerate}
 This can be written as
\begin{eqnarray}
Z_{G}(\Delta,\delta) =\sum_{\{\vu_i,\vv_a\}\in C(\delta)}\prod_{(i,a)\in E}
\ind(|\vu_{i}\cdot\vv_a-\vu^0_{i}\cdot\vv^0_a|\le \Delta), \nonumber
\end{eqnarray}
where $C(\delta)$ is the set of vectors that satisfy condition $(2)$ above.
We further define the set of \emph{typical instances} $(\M,E)$, 
${\sf Typ}(\gamma)$ through the following conditions:
\begin{enumerate}
\item Let $\theta_\U(\,\cdot\,)$ be the type of factor $\U$,
namely $n\theta_{\U}(\vu)$ is the number of row indices $i\in R$ 
such that $\vu_i=\vu$. Then for $(\M,E)\in \Typ(\gamma)$,
we have $D(\theta_\U||p_0)\le \gamma$.
\item Analogously, for the type of factor $\V$ we require
$D(\theta_\V||q_0)\le \gamma$.
\item Finally, let $\theta_E(\,\cdot\,,\,\cdot\, )$ be the edge type,
i.e. $n\eps\theta_E(\vu,\vv)$ is the number of edges $(i,a)\in E$ 
such that $\vu_i=\vu$ and $\vv_a=\vv$.  We then require 
$D(\theta_\V||p_0\cdot q_0)\le \gamma$ (where $p_0\cdot q_0$ is the product
distribution on $\vu$, $\vv$).
\end{enumerate}
By standard arguments \cite{CoverThomas} we have 
$\prob\{\Typ(\gamma)\}\to 1$ for any positive $\gamma$ as $n\to\infty$.
We then define
\begin{eqnarray}
\Zh_G(\Delta,\delta) \equiv Z_G(\Delta,\delta)\, \ind\big( (\M,E)\in\Typ(\gamma)\big)\, .
\end{eqnarray}

According to lemma \ref{NumberofSolutions}, the expectation of $\Zh_{G}(\Delta,\delta)$
vanishes as $n$ tends to infinity for $\delta>\odelta(\eps,\alpha,\Delta)$. 
Since  $\prob\{\Typ(\gamma)\}\to 1$ and using
Markov inequality, this implies that 
$\displaystyle\lim_{n\rightarrow\infty}\prob\{Z_G(\Delta,\delta)>0\} = 0$.
In conclusion, any 
matrix $\hM$ that satisfies $|\M_{ia}-\hM_{ia}|\le \Delta$ for all
$(i,a)\in E$ 
results in a distance metric smaller than 
$\odelta(\eps,\alpha,\Delta)$ with high 
probability, as $n$ tends to infinity.
\end{proof}

\begin{lemma}
\label{NumberofSolutions}
For any $\delta>\odelta(\eps,\alpha,\Delta)$ there exists $\gamma>0$ such that
 $\displaystyle\lim_{n\rightarrow\infty}\E_{E,\M}\{\Zh_G(\Delta,\delta)\} = 0$. 
\end{lemma}
\begin{proof}
$Z_G(\Delta,\delta)$ is a random variable where the randomness 
comes from the matrix elements $\M_{i,a}$ and the choice of 
the sampling set $E$. Since $E$ is uniformly random, we can 
take any realization of $\M=\U\cdot\V$ from the typical set 
according to iid $p_0$ and iid $q_0$. Given one such realization of 
$\U=(\vu_1^0,\dots,\vu_n^0)$ and $\V=(\vv_1^0,\dots,\vv_m^0)$, 
go through all the estimations $\hM=\widehat\U\cdot\widehat\V$, where 
$\widehat\U=(\vu_1,\dots,\vu_n)$ and $\widehat\V=(\vv_1,\dots,\vv_m)$.
Now group the set of assignments $\widehat\U$ and $\widehat\V$ that 
have the same empirical distribution, and let $p(\vu,\vu^0)$
and $q(\vv,\vv^0)$ denote the joint distribution.
Then, the number of different assignments with same empirical 
distribution $(p,q)$ is $e^{n\{H(p)-H(p_0)\}+m\{H(q)-H(q_0)\}}$.
For each distribution pair $(p,q)$ that satisfy condition (2)  above, 
we fix the factors $\widehat\U$ and $\widehat\V$ and  compute
the probability that they satisfies condition (1).
Denoting by $\E'_{E,\M}\{\cdots\} =\E_{E,\M}\{\cdots\ind((E,\M)\in\Typ(\gamma))\} $
the expectation restricted to $(E,\M)\in\Typ(\gamma)$, we have
\begin{align}
&\E'_{E,\M}\{Z_G(\Delta,\delta)\} \nonumber\\
&= \E'_{E,\M}\left\{\sum_{\{\vu_i,\vv_a\}\in C(\delta)}\prod_{(i,a)\in E}
                     \ind(|\vu_{i}\cdot\vv_a-\vu^0_{i}\cdot\vv^0_a|\le \Delta)\right\} \nonumber\\
&\stackrel{\cdot}{=} \sum_{\substack{p\in\cD(p_0),q\in\cD(q_0)\\d(p,q)\geq \delta}}
                          e^{nH(p|p_0)+mH(q|q_0)} \cdot\nonumber\\
&\quad\quad\quad          \E_{E}'\left\{\prod_{(i,a)\in E}\ind(|\vu_{i}\cdot\vv_a-\vu^0_{i}\cdot\vv^0_a|\le \Delta)\right\}
                          \nonumber
\end{align}
To compute the expectation in the last inequality, we 
look at a typical realization of $E$ and  partition it 
into subsets $\{E_{\vu^0,\vv^0}\},$ for 
$(\vu^0,\vv^0)\in (A_N)^r\times (A_N)^r$, defined as follows.
$(i,a) \in E$ is in $E_{\vu^0,\vv^0}$ if $\vu_i^0=\vu^0$ and $\vv_a^0=\vv^0$.
By definition $|E_{\vu^0,\vv^0}|= n\eps\theta_E(\vu_0,\vv_0)$.
Further $E_{\vu^0,\vv^0}$ is uniformly random given its size.
Within the typical set $\Typ(\gamma)$, $\theta_E(\vu_0,\vv_0)$ is close to 
$p_0(\vu^0) q_0(\vv^0)$. We thus get
\begin{align}
&\E_{E}'\left\{\prod_{(i,a)\in E}\ind(|\vu_{i}\cdot\vv_a-\vu^0_{i}\cdot\vv^0_a|\le \Delta)\right\}\nonumber\\
&\stackrel{\cdot}{=} \prod_{\vu^0,\vv^0}\E_{E_{\vu^0,\vv^0}}\left\{\prod_{(i,a)\in E_{\vu^0,\vv^0}}
                     \ind(|\vu_{i}\cdot\vv_a-\vu^0_{i}\cdot\vv^0_a|\le \Delta)\right\}\nonumber\\
&\ed \prod_{\vu^0,\vv^0}\prob\left\{|\vu_{i}\cdot\vv_a-\vu^0_{i}\cdot\vv^0_a|\le \Delta\;\big|\;
                      \vu^0,\vv^0\right\}^{n\eps \theta_E(\vu^0,\vv^0)} .\nonumber
\end{align}
Finally, we get,
\begin{align}
&\E_{E,\M}'\{Z_G(\Delta,\delta)\} \le e^{n\kappa(\gamma)}  \sum_{\substack{p\in\cD(p_0), q\in\cD(q_0)\\ d(p,q)\geq \delta}}
                      e^{n \phi_\Delta(p,q)} \label{lemma_conclusion}~.
\end{align}
where $\kappa(\gamma)\to 0$ as $\gamma\to 0$.
For $(p,q)$ that satisfies $d(p,q)>\odelta(\eps,\alpha,\Delta)$, 
we know that $\phi_\Delta(p,q)<0$ by definition. 
Hence, for $\gamma$ small enough,
$\delta>\odelta(\eps,\alpha)$ is a sufficient condition for 
$\displaystyle\lim_{n\rightarrow\infty}\E_{E,\M}\{\Zh_G(\Delta,\delta)\} = 0$.
\end{proof}

%

%
%
\subsection{General distributions via quantization}

 Above tighter upper bound can be generalized to matrices in theorem \ref{thm:MainUB}
via quantization argument. 
In this section, we're interested in 
recovering a continuous real valued matrix $\M$ from samples of its entries. 
First, we estimate it using factors $\widehat\U_{i,k}$, $\widehat\V_{k,a}$ 
supported in the continuous alphabet. 
Then, the distortion is bounded using the upper bound from section \ref{subsec:discrete}
via quantization.

\begin{propo}
Let $\Delta\ge 0$ and $\M$ be a random rank-$r$ matrix
with factors supported in continuous bounded alphabet $A_c$. 
Let $A_\delta$ be discrete quantized alphabet of $A_c$, 
with maximum quantization error less than 
$\delta/2$. $\hM$ is the rank-$r$ estimation with factors supported in $A_c$.
Then, with high probability, any matrix $\hM$
that satisfies $|\M_{ia}-\hM_{ia}|\le \Delta$ for all
$(i,a)\in E$ also satisfies $D(\M,\hM)\le \odelta(\eps,\alpha,\Delta+2err(\delta)) + 2err(\delta) 
+o_n(1)$, where $\odelta(\eps,\alpha,\Delta)$ is defined as in Eq.~(\ref{eq:Upperbound}) and 
$err(\delta)$ is the quantization error which only depends on $\delta$.
\end{propo}
\begin{proof}
Let $\M^\delta$ be the quantized version of the original matrix $\M$, 
which is defined as follows.
Define $\vu_i^\delta\in (A_\delta)^r$ and $\vv_a^\delta \in (A_\delta)^r$ to be the 
quantized version of $\vu_i$ and $\vv_a$ respectively, where $\vu_i$ is the $i$-th
row of $\U$ and $\vv_a$ is the $a$-th column $\V$. 
Then, $\M^\delta$ is defined as,
\begin{eqnarray}
\M^\delta_{i,a} = \vu^\delta_i\cdot \vv^\delta_a\, . \nonumber
\end{eqnarray}
Note that $\M^\delta_{i,a}$ satisfies $|\M_{i,a}-\M^\delta_{i,a}| \leq err(\delta)$.
Analogously, define $\hM^\delta$ to be the quantized 
version of the estimated matrix $\hM$. Then, the $\M^\delta$ and $\hM^\delta$ satisfy 
$|\hM^\delta_{i,a}-\M^\delta_{i,a}|\le \Delta+2err(\delta)$ for all $(i,a)\in E$. 

Let $\odelta(\epsilon,\alpha,\Delta)$ be the upper bound 
in proposition \ref{propo:discrete}. Then, 
the distortion is bounded with high probability by
\begin{eqnarray}
D(\M,\hM) &\leq& D(\M,\M^\delta) + D(M^\delta,\hM^\delta) + D(\hM^\delta,\hM) \nonumber \\
          &\leq& \odelta(\epsilon,\alpha,\Delta+2err(\delta)) + 2err(\delta)~ \label{ContinuousUpperbound}.
\end{eqnarray}
Note that twice the quantization error is added to $\Delta$ since 
now we only have $|\hM^\delta_{i,a}-\M^\delta_{i,a}|\le \Delta+2err(\delta)$ for all $(i,a)\in E$.
\end{proof}
%
%
%
\subsection{Simplified bound}

The (tighter) upper bound in proposition \ref{propo:discrete} is not easily computed.
To get a bound that can be analyzed, we relax the constraint $\phi_\Delta\geq 0$ 
and get a relaxed or simplified upper bound on $\odelta(\eps,\alpha,\Delta)$.
Furthermore, this simplified upper bound is used to prove theorem \ref{thm:MainUB}.
\begin{propo}
\label{propo:simplebound}
For all $\eps\geq 0$, $\alpha\geq 0$ and $\Delta\geq 0$, we have 
\begin{align}
&\odelta(\eps,\alpha,\Delta) \leq \nonumber\\
&\left\{\dmax^2-(\dmax^2-\Delta^2)
\exp\left(-\frac{\overline{H}(p|p_0)+\alpha\overline{H}(q|q_0)}{\eps} \right)\right\}^{1/2}~, \nonumber
\end{align}
where 
$\odelta(\eps,\alpha,\Delta)$ is defined as in proposition \ref{propo:discrete},
$\overline{H}(p|p_0)=\displaystyle\max_{p\in\cD(p_0)}\{H(p)\}-H(p_0)$, 
 $\overline{H}(q|q_0)=\displaystyle\max_{q\in\cD(q_0)}\{H(q)\}-H(q_0)$, 
and $\dmax = max\{|\vu\cdot\vv-\vu^0\cdot\vv^0|\}$.
\end{propo}
\begin{proof}
Define the upper bound $\sdelta(\eps,\alpha,\Delta)$ as
\begin{eqnarray}
\sdelta(\eps,\alpha,\Delta) = \!\!\sup_{ \substack{p\in\cD(p_0)\\q\in\cD(q_0)}}\!\!\!
\left\{\;d(p,q)\,:\;\; \phi^u_{\Delta}
(p,q)\;\ge 0\right\}\, ,\label{eq:SimplifiedUpperbound}
\end{eqnarray}
where $\cD(p_0)$, $\cD(p_0)$ and $d(p,q)$ are defined
 in Eq.~(\ref{eq:Upperbound}). The only 
difference is the relaxed constraint function $\phi^u_{\Delta}$, defined as
\begin{align}
\phi^u_\Delta(p,q) &\equiv \overline{H}(p|p_0)+\alpha \overline{H}(q|q_0)+
\eps\, \log\left(
\frac{\dmax^2-d(p,q)^2}{\dmax^2-\Delta^2}\right)\, .\nonumber
\end{align}
By Jensen's and Markov inequality, $\phi^u_\Delta(p,q)$ is 
larger than $\phi_\Delta(p,q)$.
This implies that the supremum in the simplified upper bound is taken over 
a larger set of distributions than the tighter upper bound, hence 
we have $\odelta(\eps,\alpha,\Delta) \leq \sdelta(\eps,\alpha,\Delta)$.
And after some computation, it's easy to show that $\sdelta(\eps,\alpha,\Delta) = 
\left\{\dmax^2-(\dmax^2-\Delta^2)
\exp\left(-\frac{1}{\eps}\left[\overline{H}(p|p_0)+
\alpha\overline{H}(q|q_0)\right]\right)\right\}^{1/2}$~,
which concludes the proof.
\end{proof}

This simplified upper bound can be generalized, 
in the same manner, to the continuous support case. 
The following example illustrates this generalization 
and introduces bounds necessary in the proof of theorem \ref{thm:MainUB}.

For the original matrix $\M = \U\cdot \V$, assume the distributions of 
$\U_{i,k}$ and $\V_{k,a}$ to have support in 
$A_\delta=\{-1,-1+\delta,\dots,1-\delta,1\}$. Also, the factors of
the rank-$r$ solution $\hM$ are supported on the same discrete set.
Then, the simplified upper bound is given by
\begin{align}
&\sdelta(\eps,\alpha,\Delta) = \nonumber\\
&\left(\Delta^2+(4r^2-\Delta^2)
\left(1-\exp\left\{-\frac{\log{N}}{\te}\right\}\right)\right)^{1/2} \, ,\nonumber
\end{align}
where $N=|A_\delta|$ and $\te \equiv \eps/(1+\alpha)r$.
Note that $\displaystyle\lim_{\eps\rightarrow\infty} \sdelta(\eps,\alpha,\Delta)$ = $\Delta$,
which means that we cannot get RMSE smaller than $\Delta$.

The maximum quantization error associated with $M_{i,a}$ is 
$r(\delta-\delta^2/4)$, which happens when all the entries of 
$\vu_i^0$ and $\vv_a^0$ are $1-\delta/2$ and quantized to $1$.
For simplicity, $err(\delta)=r\delta$ is used.
Combined with Eq.~(\ref{ContinuousUpperbound}), 
we have a simple analytical upper bound on the distortion when 
the original matrix and the estimation 
have continuous support $[-1,1]$.

\begin{figure}[h]
\center{\includegraphics[width=8.cm]{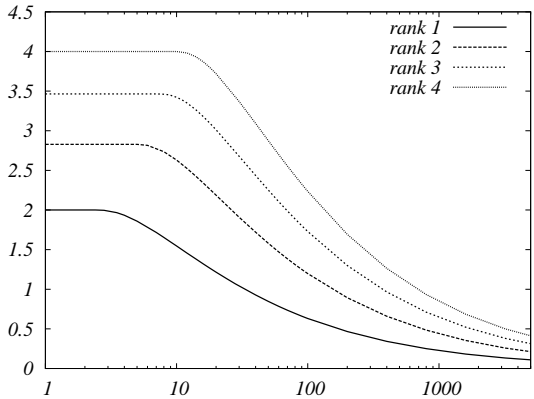}}
\caption{{\small 
The upper bound in Eq.~(\ref{ContinuousUpperbound}) with simplified upper bound $\odelta^u(\eps,\alpha,\Delta)$, for 
$\alpha=1$ and $\Delta=0$ and a few values of the rank $r$.}}
\label{fig:Contbounds}
\end{figure}
%

%
%
%
%
%
%
%
\begin{proof}[Proof of Theorem \ref{thm:MainUB}]
From the example above, we can compute the simplified upper bound directly 
to bound the distortion.
\begin{align}
&D(\M,\hM) \nonumber\\
&\leq \left\{4r^2\!-\!(4r^2\!-\!(\Delta\!+\!2r\delta)^2)
      \left(\exp{\left(\frac{-\log{N}}{\te}\right)}\right)\right\}^{1/2} 
\!\!\!\!+\! 2r\delta \nonumber\\
&\leq \left\{(\Delta+2r\delta)^2+4r^2
      \left(1-\exp\left(-\frac{\log{N}}{\te}\right)\right)\right\}^{\frac{1}{2}} + 2r\delta \nonumber\\
&\leq \Delta+4r\delta+2r\left(1-\exp\left(-\frac{\log{N}}{\te}\right)\right)^\frac{1}{2} \nonumber\\
&\leq \Delta+4r\delta+2r\left(\frac{\log{N}}{\te}\right)^\frac{1}{2}~. \nonumber
\end{align}
Remember N is defined as the alphabet size $|A_\delta|$, 
where the discrete alphabet $A_\delta=\{-1,-1+\delta,\cdots,1-\delta,1\}$ is used. 
Fixing $\delta=\frac{2}{N-1}$, we can minimize the right hand side 
of the last inequality with respect to the alphabet size $N$. 
Since the exact minimizer cannot be represented in a closed form, 
we use instead an approximate minimizer $N=\left\lceil{4\sqrt{\te}}\right\rceil+1$, 
which results in
\begin{align}
&D(\M,\hM) \nonumber\\
&\leq \Delta+2r\left\{\frac{4}{\left\lceil{4\sqrt{\te}}\right\rceil}+
      \left(\frac{\log\left(\left\lceil{4\sqrt{\te}}\right\rceil+1\right)}{\te}\right)^\frac{1}{2}\right\} \nonumber\\
&\leq \Delta+\frac{2r}{\sqrt{\te}}\left\{1+
      \left({\log\left(\left\lceil{4\sqrt{\te}}\right\rceil+1\right)}\right)^\frac{1}{2}\right\} \nonumber\\
&\leq \Delta+\frac{2r}{\sqrt{\te}}\log\left({10\te}\right)~, \label{proof_theorem}
\end{align}
where the last inequality in (\ref{proof_theorem}) is true for $\te>1.5$.
This is practical since we are typically interested in the region where 
$\frac{\log(10\te)}{\sqrt{\te}} \leq 1$.

\end{proof}
%
%
\section{Lower Bound}
\label{sec:LB}

When the number of observed elements is smaller than $\Theta(n)$,
high distortion is inevitable. In this section 
we derive a quantitative lower bound which 
supports this observation.
\begin{propo}
\label{propo:lowerbound}
 Let $\M = \U\cdot \V$ be a random rank-$r$ matrix with $n$ rows and
$n\alpha$ columns and assume the distributions of $\U_{i,k}$ and $\V_{k,a}$ 
to have support in $[-1,1]$, and $E$ a random subset of $n\eps$ 
row-column pairs.
Then, with high probability, any rank-$r$ matrix $\hM$ such that 
$|\M_{i,a}-\hM_{i,a}| = 0$ for all $(i,a)\in E$, also satisfies
\begin{eqnarray}
D(\M,\hM) \geq \tilde{c}\cdot e^{-\eps} ~, \label{lowerbound1}
\end{eqnarray}
where $\tilde{c}$ is a strictly positive 
constant that only depends on the rank $r$ 
and the initial distributions $p_0$ and $q_0$.
\end{propo}
\begin{proof}
 Think of the following algorithm which has clearly better performance than 
any other that satisfies the assumptions. 
Consider the bipartite graph $G = (R,C,E)$ with 
vertices corresponding to the row and columns of $\M$ and edges 
for the observed entries. For every pair of row and column indices  
$(i,a)$, $i \in R$ and $a \in C$, that is not connected by an edge, 
we do the following. If degree of $i$ ($a$) is less than $r$, 
we assume that all the neighbors of node $i$ ($a$) are known and 
make MMSE estimation of $\vu_i^0$ ($\vv_a^0$). 
If degree of $i$ ($a$) is greater than $r-1$, 
we assign the correct value of $\vu_i^0$ ($\vv_a^0$). 
With high probability the resulting RMSE is greater than 
$\underline\delta(\eps,\alpha)$ as defined below.
\begin{align}
\underline\delta(\eps,\alpha) = \sqrt{(1-(1-\xi)(1-\zeta))\tilde{c}}~, \label{lowerbound2}
\end{align}
where $\xi=
\prob\{degree(i)<r\}=\displaystyle\sum_{k=0}^{r-1}{\frac{\eps^{-k}}{k!}e^{-\eps}}$~,
$\zeta=
\prob\{degree(a)<r\}=\displaystyle\sum_{k=0}^{r-1}{\frac{(\eps/\alpha)^{-k}}{k!}e^{-\eps/\alpha}}$
and $\tilde{c}=\min\{\E\{\vu_i^0\cdot(\vv_a^0-\vv_a')\},\E\{(\vu_i^0-\vu_i')\cdot\vv_a^0\}\}$. 
Here, $\vu_i'$ and $\vv_a'$ represent the MMSE estimate of $\vu_i^0$ and $\vv_a^0$ respectively,
assuming that $r-1$ neighbors and corresponding edges are known. 

Without loss of generality, assume $\alpha \geq 1$. 
Then, we can simplify above bound to get, Eq.~(\ref{lowerbound1})
\end{proof}
\begin{figure}[h]
\center{\includegraphics[width=8.cm]{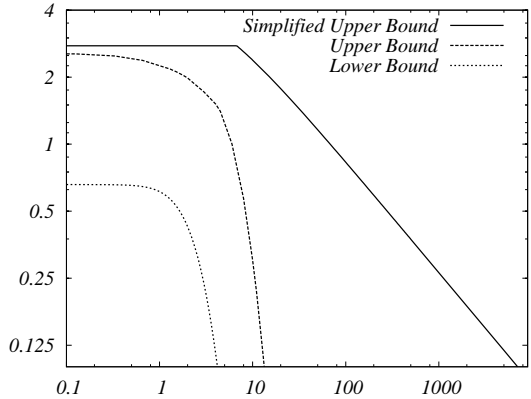}}
\caption{{\small The upper bound $\odelta(\eps,\alpha,\Delta)$, 
the simplified upper bound $\sdelta(\eps,\alpha,\Delta)$ and 
the lower bound $\underline\delta(\eps,\alpha)$ 
for rank $r = 2$, $\alpha=1$, $\Delta=0$.
Here the factors $\U_{ik}$, $\V_{ka}$ take values in  $\{-1,0,1\}$.}}
\label{fig:bounds}
\end{figure}
%
%
%
\section{Efficient matrix completion}
\label{sec:Numerical}

In the previous sections we proved that $O(n)$ random entries 
determine a random low rank matrix within an arbitrarily small RMSE.
How hard is it to find such a matrix? In this section we present a 
numerical investigation using a low complexity stochastic local
search algorithm that we call WalkRank.

WalkRank is inspired by successful local search algorithms for
constraint satisfaction problem, such as WalkSAT \cite{Selman}.
It is particularly suited to low-rank matrices whose factors 
$\U_{i,k}$, $\V_{k,a}$ take values in a finite set $A_N$.
The algorithm tries to find assignments of the vectors
$\{\vu_1,\dots,\vu_n\}$, and $\{\vv_1,\dots,\vv_m\}$ that minimize
the cost function
\begin{eqnarray}
\cC(\{\vu_i,\vv_a\}) = \sum_{(i,a)\in E}\, 
\ind(|\vu_i\cdot\vu_a-\M_{ia}|>\Delta)\, ,
\end{eqnarray}
which counts the number of observations $\M_{ia}$ that are not described 
by the current assignment.

The algorithm initializes the vectors $\{\vu_i\}$, $\{\vv_a\}$ to
random iid values and then alternates between two type of moves.
The first are greedy moves, described here in the case of $\U$
factors.
\vspace{0.3cm}

{\normalsize
\begin{tabular}{ll}
\hline
\multicolumn{2}{l}{Greedy move, $\U$ factors}\\
\hline
1: & Sample a column index $i\in C$ uniformly; \\
2: & Find  $\vu^{\mbox{\tiny new}}_i$ that minimizes
$\cC(\{\vu_i,\vv_a\})$ over $\vu_i$;\\
3: & Set $\vu_i\leftarrow \vu^{\mbox{\tiny new}}_i$\\
\hline
\end{tabular}}
\vspace{0.3cm}

Greedy moves for $\V$ factors are defined analogously.

The second type of move potentially increases the cost function.
\vspace{0.3cm}

{\normalsize
\begin{tabular}{ll}
\hline
\multicolumn{2}{l}{Walk move}\\
\hline
1: & Sample  $(i,a)\in E$ s.t. $|\vu_i\cdot\vv_a-\M_{ia}|>\Delta$; 
\\
2: & Find $\vu^{\mbox{\tiny new}}_i\cdot\vv^{\mbox{\tiny new}}_a$ such that $|\vu^{\mbox{\tiny new}}_i\cdot\vv^{\mbox{\tiny new}}_a-\M_{ia}|\le\Delta$\\
3: & Set $\vu_i\leftarrow \vu^{\mbox{\tiny new}}_i$, and $\vv_a\leftarrow \vv^{\mbox{\tiny new}}_a$\\
\hline
\end{tabular}}
\vspace{0.3cm}

WalkRank recursively executes one of these moves, choosing a walk
move with probability $\rho$, and a greedy one with probability 
$1-\rho$. The parameter $\rho$ can be optimized over, and we found 
$\rho\approx 0.1$ to be a reasonable choice.
\begin{figure}[h]
\center{\includegraphics[width=8.cm]{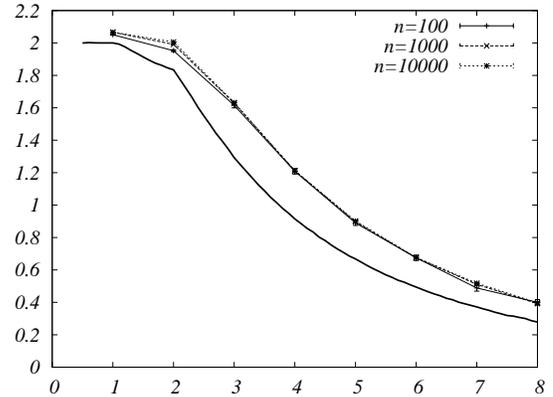}}
\caption{{\small Performances of the WalkRank algorithm on random 
rank 2 matrices. The bold line is a lower bound on the distortion
obtained by the maximum likelihood algorithm.}}
\label{fig:Rank2Walk}
\end{figure}
\begin{figure}[h]
\center{\includegraphics[width=8.cm]{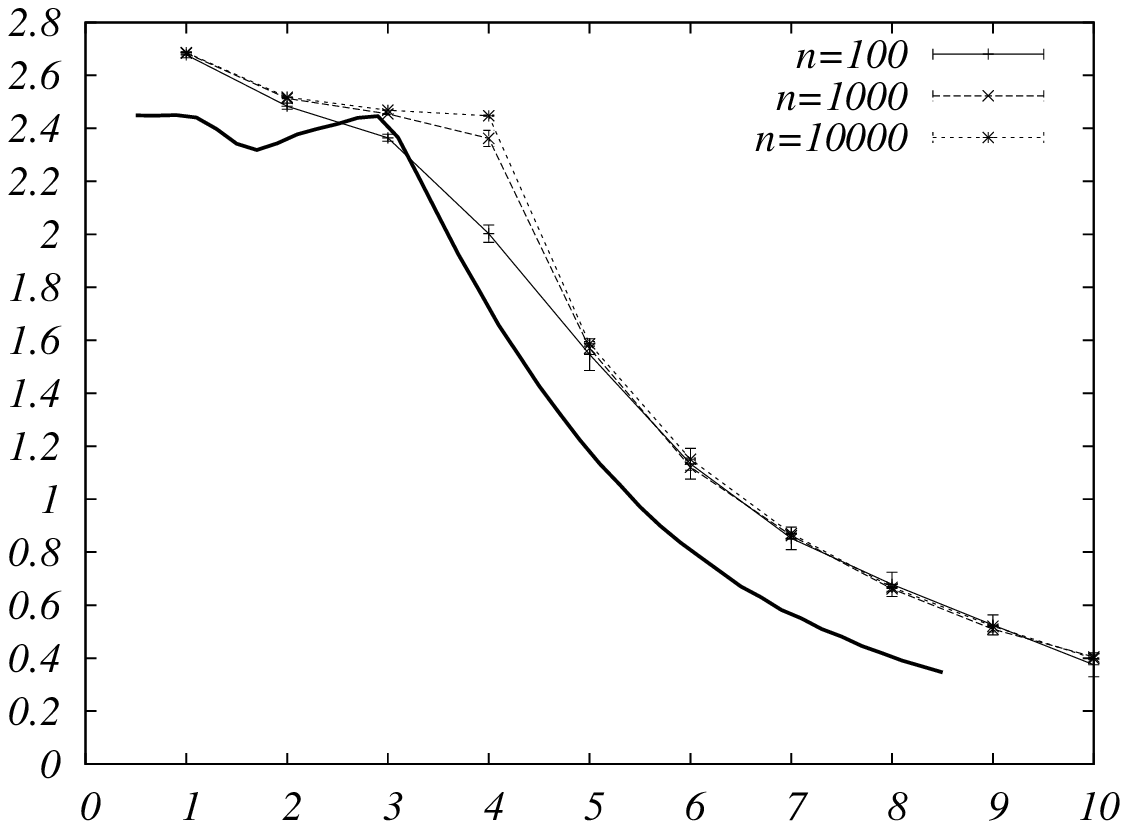}}
\caption{{\small Performances of the WalkRank algorithm on random 
rank 3 matrices.}}
\label{fig:Rank3Walk}
\end{figure}
\begin{figure}[h]
\center{\includegraphics[width=8.cm]{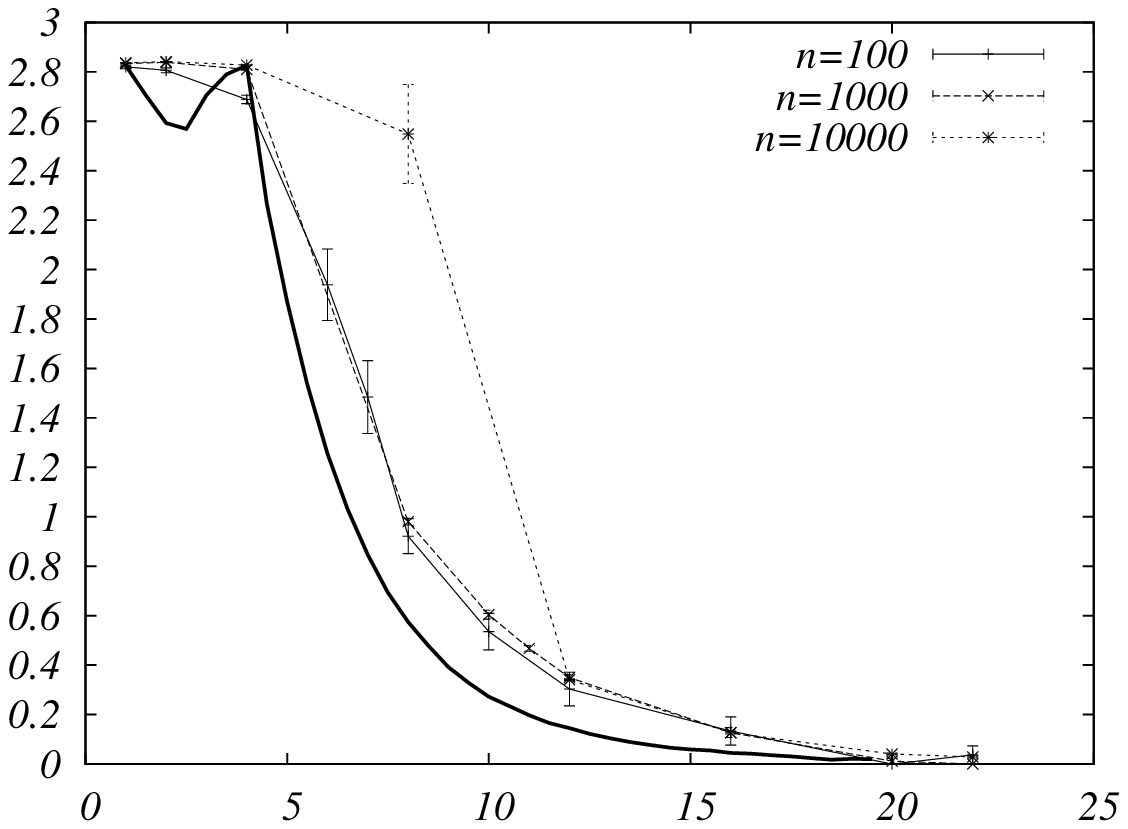}}
\caption{{\small  Performances of the WalkRank algorithm on random 
rank 4 matrices.}}
\label{fig:Rank4Walk}
\end{figure}
\begin{figure}[h]
\center{\includegraphics[width=8.cm]{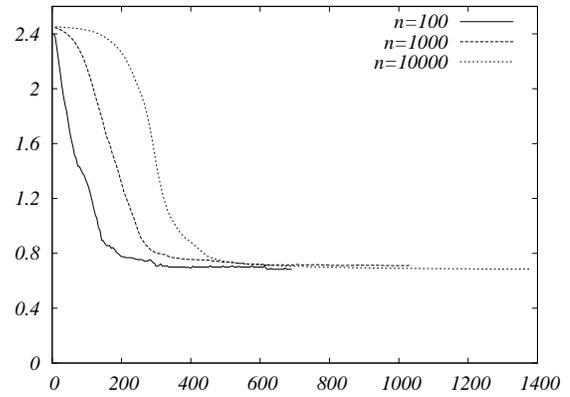}}
\caption{{\small Typical evolution of the cost function under the WalkRank 
algorithm. Here the rank is $r=3$, and $\eps=8$.}}
\label{fig:CostTime}
\end{figure}

In Figures \ref{fig:Rank2Walk} to \ref{fig:Rank4Walk}
we present the distortion achieved by the WalkRank algorithm, averaged over
$10$ instances. We used factors with entries $\U_{i,k}$, $\V_{k,a}$
uniformly distributed in $\{+1,-1\}$. It is clear that the resulting 
distortion is essentially independent of $n$ over two orders of magnitude
and decreases rapidly with $\eps$.

We compare these numerical results with an analytical lower bound on the
distortion achieved by a maximum likelihood algorithm. The latter
fills each unknown position in $\M$ with its most likely value.
While there exists no practical implementation of the maximum likelihood
rule, we can provide a sharp lower bound on its performances using 
techniques explained in \cite{Future}. It appears that,
for low values of the rank, WalkRank achieves
the same distortion as maximum likelihood, provided it is given 
one or two more entries per column/row.

The complexity of one WalkRank step is independent of the matrix size
(but grows with the rank). The results in Figures  \ref{fig:Rank2Walk} to 
\ref{fig:Rank4Walk} were obtained with a number of steps slightly superlinear 
in $n$. In Fig.~\ref{fig:CostTime} we show the evolution of the cost function
for averaged over $10$ instances for $n=10^3$ to $10^5$, $r=3$ and
$\eps=8$. The number of steps per variable required to reach the 
asymptotic value increases mildly with $n$. A reasonable conjecture
is that the number of steps scales like $n\cdot$Poly$(\log n)$.
%
%
\section{Back to the Netflix Data}
\label{sec:Netflix}

As shown in the last section, local search algorithms efficiently 
fit low rank matrices of very large dimensions, using few observations.
They therefore provide an efficient tool for checking whether a dataset is
well described by the random low rank model.

\begin{figure}[h]
\center{\includegraphics[width=7.cm]{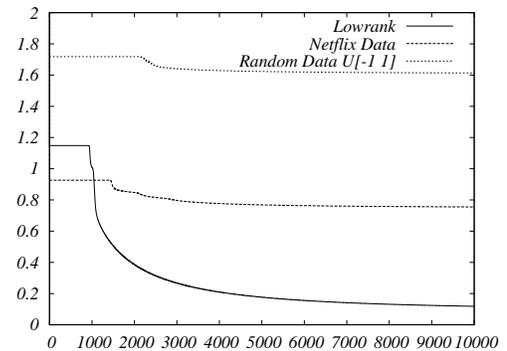}}
\center{\includegraphics[width=7.cm]{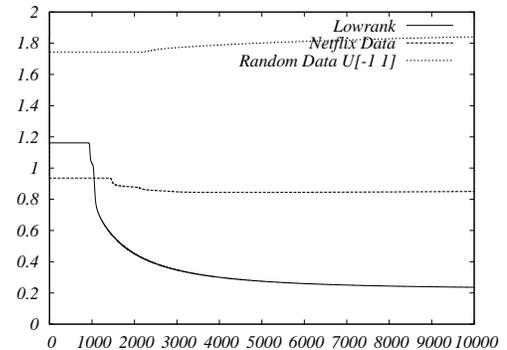}}
\caption{{\small Evolution of the fit error (top frame) and prediction
error (lower frame) for fitting three matrices with a rank $3$ model.
The curves are obtained using  coordinate descent in the factors.}}
\label{fig:Netflix3}
\end{figure}
\begin{figure}[h]
\center{\includegraphics[width=7.cm]{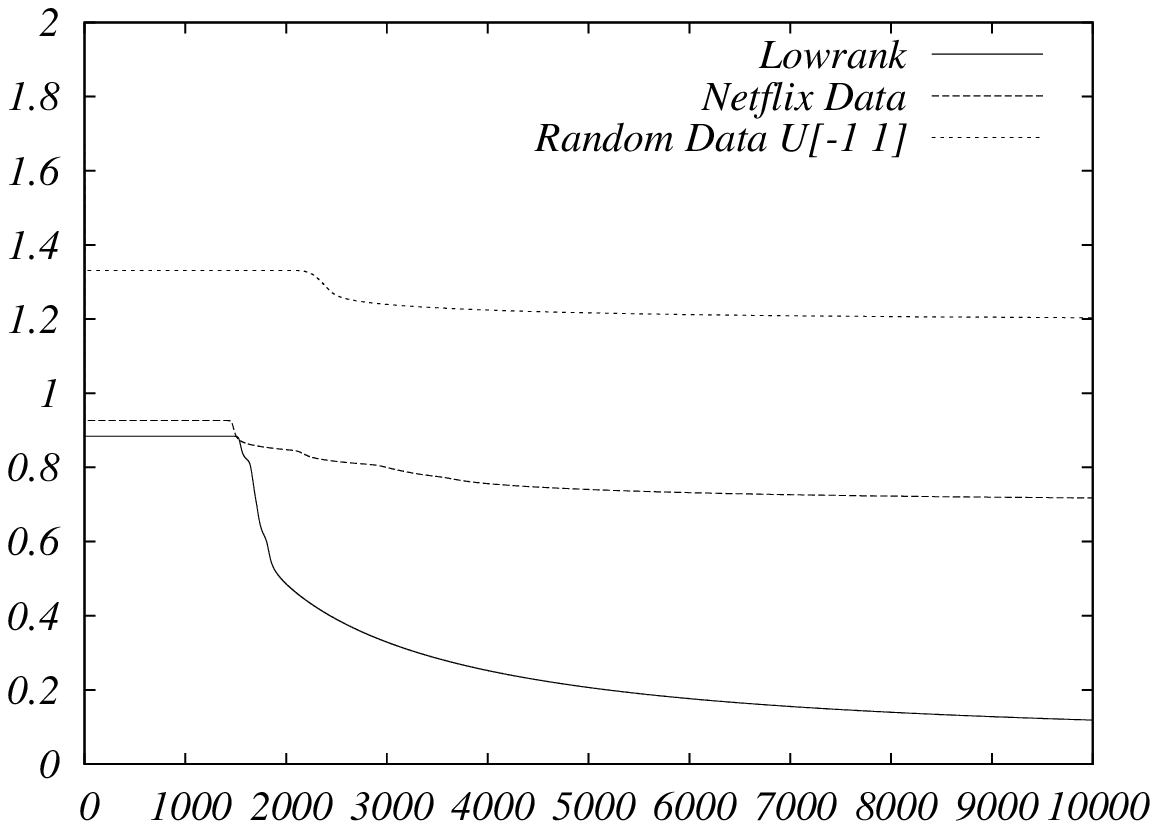}}
\center{\includegraphics[width=7.cm]{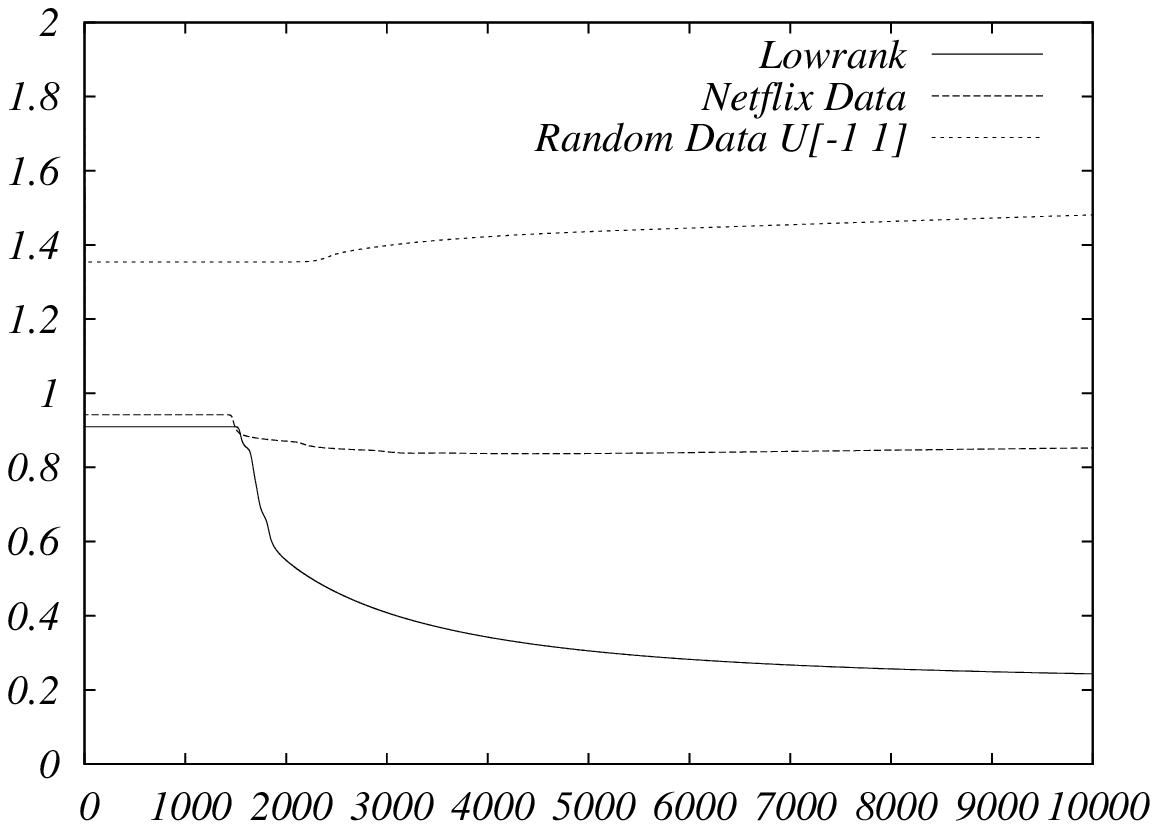}}
\caption{{\small  As in Figure \ref{fig:Netflix3}, but for a 
rank $5$ model.}}
\label{fig:Netflix5}
\end{figure}

In Figures \ref{fig:Netflix3} and \ref{fig:Netflix5} we compare the
evolution of fit and prediction error for three matrices with 
$n=m=5\cdot 10^3$:
\begin{enumerate}
\item[1.] A submatrix of the Netflix dataset given by the first 
$5\cdot 10^3$ movies and customers.
\item[2.] A matrix with the same subset $E$ of revealed entries, each of them
chosen uniformly at random in $[-1,+1]$.
\item[3.] A random rank-$3$ matrix (for Fig.~\ref{fig:Netflix3}) or 
rank-$5$ matrix (for Fig.~\ref{fig:Netflix5}), with set of revealed
entries as above.
\end{enumerate}
The fit error is defined by restricting the average in Eq.~(\ref{eq:RMSEDef})
to $(i,a)\in E$. The prediction error is instead obtained by 
averaging over $(i,a)\not\in E$. In the case of the Netflix matrix
the latter was estimated by hiding $10^3$ entries from the dataset,
and averaging over those.

We used a coordinate descent algorithm in the factors $\{\vu_i\}$,
$\{\vv_a\}$, with regularized cost function given by 
Eq.~(\ref{eq:Regularized}). In agreement with the results of previous 
sections, random low rank matrices are efficiently fitted  
with small fitting \emph{and} prediction error. The difference
with iid entries is striking. The fit error decreases only slowly over time,
while the prediction error actually increases. As expected, 
revealed entries do not provide any information on the hidden ones. 
Netflix data lie somewhat in between: both fit and prediction error decrease 
over time, albeit not as sharply as for genuine low rank matrices.
%
%
\section*{Acknowledgements}

Andrea Montanari is partially supported by a Terman fellowship
and by the NSF CAREER grant 0743978.

%
%
\bibliographystyle{IEEEtran}

\end{document}